# A Regression-based Voltage Estimation Method for Distribution Volt-Var Control with Limited Data

Catie McEntee
FREEDM System Center
North Carolina State University
Raleigh, NC, USA
cmmcente@ncsu.edu

Ning Lu
FREEDM Systems Center
North Carolina State University
Raleigh, NC
nlu2@ncsu.edu

David Lubkeman
FREEDM Systems Center
North Carolina State University
Raleigh, NC
dllubkem@ncsu.edu

*Abstract*— **This paper presents a regression-based method for estimating voltages and voltage sensitivities for volt-var control on distribution circuits with limited data. The estimator uses power flow results for representative load and PV output scenarios as training data. Using linear regressions on power flow results, the voltages at critical nodes are calculated online based on power measurements at the feeder head and at each PV plant. The voltage sensitivity to changes in reactive power injection by each PV plant is also found online using regressions on power flow results. The estimator thus provides the estimated critical voltages and their sensitivities to each possible control action. The estimator is tested in conjunction with a volt-var optimization on real, unbalanced rural distribution feeder. The optimal control actions and voltage results using the estimator are compared to the optimal results assuming full visibility of the distribution system. Results show that the estimator can estimate voltages and sensitivities with adequate accuracy for successful centralized volt-var control.**

## I. INTRODUCTION

Inverter-based distributed energy resources (DERs), such as solar photovoltaic (PV) and battery energy storage systems, have superb controllability in terms of response accuracy and speed when regulating real and reactive power outputs. However, there are also many operational uncertainties in DER operation. For example, fast fluctuations in solar irradiances can cause voltage issues (e.g., frequent over- and under- voltage events) in distribution circuits with high PV penetrations [1].

In recent years, coordinative volt-var control algorithms are proposed by researchers [2]-[8] for controlling smart inverters in DERs in coordination with voltage regulators (VRs) to replace the conventional volt-var control algorithms, which focus mainly on controlling utility-owned VR devices using local measurements. Although the performance of those algorithms are satisfactory, there are two key drawbacks.

The first drawback is an increasing requirement on measurement data and communication bandwidth. Voltage sensitivity calculations [2] and centralized volt-var algorithms [3]-[6] rely on full visibility of nodal loads, voltages, PV outputs, and circuit parameters. However, distribution networks in reality have low visibility with only a few sensors located at the feeder head and possibly at controllable devices such as reclosers, PV plants, and voltage regulators. Visibility can be achieved through state estimation algorithms, however conventional state estimations require online weighted-least-squares or similar optimization, which requires an online circuit model and can be computationally intensive, and often rely on pseudomeasurements which can introduce significant errors [9].

Data-driven, regression-based methods for estimating system operating conditions have been proposed in [8],[10]-[12]. In [10], Weckx *et al.* proposed a method for calculating voltage sensitivity factors using historical smart meter data without relying on a circuit model. This approach requires full coverage with smart meter data for each feeder over at least one year, which may not be available. In [11], Deboever *et al.* proposed a multiple linear regression method trained on just nine power flow solutions to estimate the voltages throughout a distribution feeder with one large PV plant for any load and PV output combination. However, the authors didn't consider the impact of load diversity on voltage profiles. In [8] and [12], Zhang *et al.* and Rigoni *et al.* respectively proposed polynomial regressions on power flow results to predict voltages and sensitivity factors in a distributed manner based on measurements at each DER location. These distributed approaches require controllers at each DER location which can operate based on trained regressions. Regression-based approaches for centralized volt-var control schemes, where control decisions are made by a central controller with wider knowledge of the entire feeder state, have not yet been explored. Therefore, a computationally efficient and accurate voltage and voltage sensitivity estimation method that requires few real-time input data is needed for accurate, efficient, and cost-effective centralized volt-var control (CVVC).

The second drawback relates to the implementation of VR control actions. The volt-var control algorithms in [5]-[7] calculate optimal tap changes and thus require implementation of specific tap positions at each VR. However, conventional regulator controllers are not designed to implement specific tap positions; rather, a voltage set point, together with a predetermined voltage dead band, are used to determine the tap changer operation [1]. Thus, instead of sending a tap position, the VR control signal needs to be translated to a voltage set point.

Funded by the U.S. Department of Energy Advanced Research Projects Agency – Energy

XXX-X-XXXX-XXXX-X/XX/$XX.00 ©20XX IEEE

To resolve those two issues, in this paper, we propose *a machine learning based approach* for estimating voltages and voltage sensitivities online using measurements from a small number of sensors and *a mapping algorithm* for translating VR tap positions into voltage set points so the CVVC can be used with existing regulator controllers in the field. To prepare the offline training data for the regression-based estimator, a novel representative-scenario selection method is developed. The regression-based estimator predicts the node voltages at critical nodes using power measurements only at the feeder head and at the PV plant locations and estimates voltage sensitivities to control actions, including reactive power changes from DER inverters and VR tap changes. Trained offline, the estimator is effective on circuits with multiple large PV plants and VRs and diverse load profiles.

When used for online applications, the regression-based voltage and sensitivity estimators only require active and reactive power measurements at the feeder head and large PV plants. This will substantially reduce the need for extensive monitoring and communication infrastructure and reduces the computations needed to calculate the voltage sensitivities online. Limited online voltage measurements can optionally be used to improve the accuracy of the voltage estimator in the presence of load diversity and modeling errors.

To demonstrate the effectiveness of the proposed algorithm, we compare the regression-based method with the power flow based CVVC on a real, unbalanced distribution feeder. Results show that the regression-based approach achieves adequate accuracy in CVVC with less communication needs and shorter computing time.

The main contributions of this paper are summarized as follows: 1) a novel representative training scenario selection considering load diversity; 2) a critical-nodes based constrained volt-var optimization algorithm for reducing the computational speed, 3) a fast and accurate regression-based centralized voltage and voltage sensitivity estimation algorithm that is effective on large circuits with multiple large PV plants and VRs, 4) the online voltage estimation correction mechanism using limited voltage measurements, and 5) the tap-position-to-set-point mapping method for seamless implementation.

The rest of the paper is organized as follows. In section II, the critical node selection and voltage and sensitivity estimators are described. Section III lays out the optimization formulation. In section IV, the test feeder and simulation setup are discussed. Section V shows the simulation results and section IV summarizes the findings and future work.

## II. METHOD

The training and online application processes of the proposed regression-based volt-var controller is illustrated in Fig. 1. There are two objectives in the offline training phase: minimize the measurement and communication infrastructure needs by only collecting data from a limited number of existing sensors and accelerate the computation speed so optimization can be achieved inside each control interval.

To achieve the two objectives, it is necessary to use a regression-based estimator for predicting nodal voltage and voltage sensitivity. The estimator is trained with results from offline power flow cases using historical data. In this paper, we propose a centralized method which only requires regressions housed in one central controller. We introduce representative scenario selection and voltage correction to address load diversity and the tap-position-to-set-point mapping to implement VR actions.

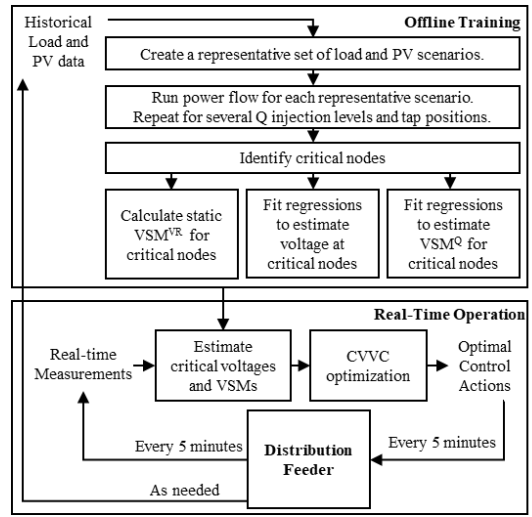

Fig. 1. Flowchart of the offline training and online operation processes

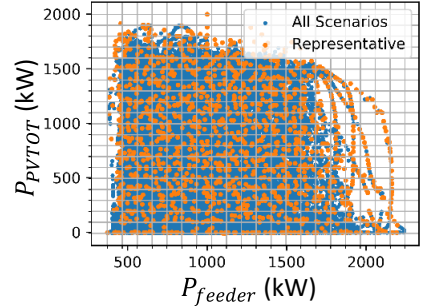

Fig. 2. Selection of typical operation scenarios

### A. Representative Scenario Selection

As a first step, collect one year of historical smart meter data for a distribution feeder. If smart meter data is not available, the disaggregation algorithm in [13] can be used to create diverse synthetic smart meter data for each load node. Then, plot the feeder head load, $P_{feeder}$, versus the total PV power output, $P_{PVTOT}$ on a scatter plot, as shown in Fig. 2. Divide the plot into blocks. For each scenario, calculate the load and PV centers, $D_L$ and $D_{PV}$ as

$$D_L = \frac{1}{\sum_{i \in N} L_i} \sum_{i \in N} D_i \times L_i \quad (1)$$

$$D_{PV} = \frac{1}{\sum_{i \in N} P_i} \sum_{i \in N} D_i \times P_i \quad (2)$$

where $D_i$ is the electrical distance of node $i$ from the feeder head; $L_i$ and $P_i$ are the load and PV injection at node $i$, respectively. Note that the load and PV centers provide a measure of how the load and PV injections are distributed throughout the feeder.

Next, choose 6 scenarios (i.e., maximum, median, and minimum load and PV centers shown by the orange dots) from each block. These scenarios will represent all the scenarios with similar $P_{feeder}$ and $P_{PVTOT}$ levels and are chosen because they represent the cases where the maximum, minimum, and median voltages are expected based on the load and PV distributions, as shown in Fig. 3. By choosing these specific scenarios from each block, the key variations of scenarios can be captured to reduce the overall bias in the training data.

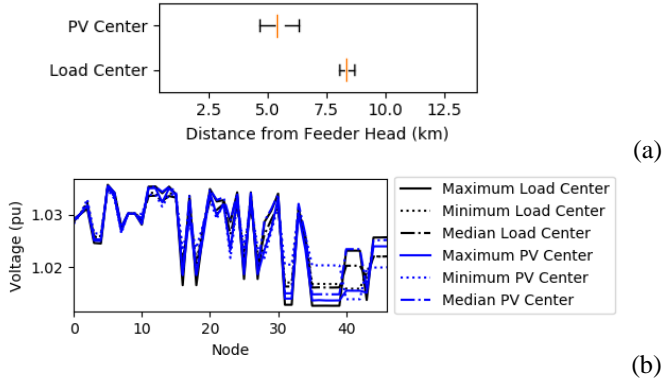

Fig. 3. (a) Minimum, maximum, and median load centers for one load/PV block and (b) resulting voltage profiles

### B. Training Data Preparation

The training data generation process is detailed in algorithm 1. For the first selected representative scenario, perturb each regulator's tap up/down and run power flow to establish the relationship between the tap changes and nodal voltage variations. Then, for each selected representative scenario, set the reactive power injection at each PV plant at different levels and run power flow for each case to establish the relationship between PV reactive power injection/absorption and nodal voltage variations.

| Algorithm 1: Training Data Generation |
|---|
| 1:     Select representative scenarios |
| 2:     Solve power flow for the first scenario (no PV reactive power injection and no VR actions) |
| 3:     **for** each voltage regulator |
| 4:       Increase the tap position by 1 |
| 5:       Record the voltage change |
| 6:       Return the tap to the initial position |
| 7:     **for** each scenario |
| 8:       **for** each PV plant |
| 9:         **for** each Q value in (-100%,-80%...100%) |
| 10:           **if** Q is within the PV excess capacity |
| 11:             Set PV reactive power to Q |
| 12:             Solve power flow |
| 13:             Record Voltages and Power measurements |
| 14:           **else** |
| 15:             Skip to next Q value |
| 16:           **end** |
| 17:         **end** |
| 18:       revert PV reactive power to 0 |
| 19:       **end** |
| 20:     **end** |

### C. Critical Nodes Selection

In practice, the CVVC optimization problem is constrained by nodes with extreme voltages. Therefore, we define critical nodes as the nodes where the highest or lowest voltages on the feeder are most likely to occur. Thus, by narrowing the consideration from all nodes to only the critical nodes, we can simplify the CVVC optimization problem significantly.

As shown in Fig. 3, the initially selected critical nodes include the feeder head node, the primary and secondary nodes of each voltage regulator and the PV plant locations where voltage rise can cause extreme voltages. Then, after each scenario run, any node with the highest or lowest voltage will be added to the critical node list. Note that such nodes usually occur near the ends of the feeder.

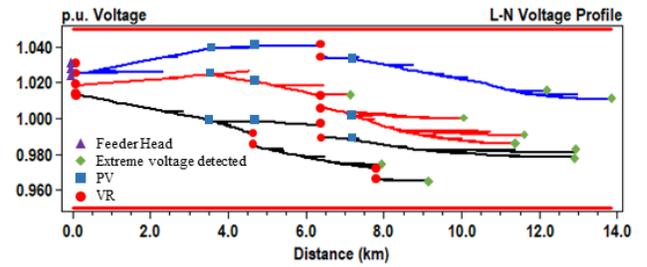

Fig. 3. Voltage profile with critical nodes

Estimating voltage only at the critical nodes ensures that the online optimization will be able to address the most extreme voltage violations without considering voltage constraints at every node in a large system. This can also reduce the number of regressions that are needed as part of the estimator, as each node considered requires one regression for voltage estimation and one regression for voltage sensitivity estimation.

### D. Voltage Estimator and Voltage Sensitivity Estimator

Linear regression is used in the voltage estimator to calculate voltage from active and reactive power measurements on each phase at the feeder head and at each PV plant. Least squares regression is used to fit the power flow nodal voltages with respect to the power measurements. The voltage-sensitivity estimator estimates the voltage sensitivities with respect to PV reactive power injections/absorptions for each scenario using power flow results. For each power flow result, after the PV reactive power for one plant is perturbed, the voltage values are subtracted from the voltage values of the previous power flow result. The voltage sensitivity is the change in voltage divided by the change in reactive power. The voltage sensitivities are calculated for each critical node with respect to each PV plant across all power flow results.

Figure 4 shows the voltage sensitivity, $\delta^Q$, for one critical node with respect to one PV plant and how it changes with respect to various possible feeder measurements. The trends with respect to voltages at the PV node, voltages at the critical node being considered, active and reactive power at the feeder head, and active and reactive power at the PV plant are shown. The voltage sensitivity shows a clear trend with respect to the reactive power at the feeder head. Therefore, the reactive power measured at the feeder head is the sole variable used to regress the voltage sensitivity on. Based on the inverse shape of the curve, $\delta^Q$ can be fit by a curve using reactive power at the feeder head, $Q_s$, for each critical node and PV plant pair as

$$\delta^Q = a/(3000 + Q_s)^b \quad (3)$$

The resulting curve and residuals for one critical node with respect to one PV plant are shown in Fig. 5.

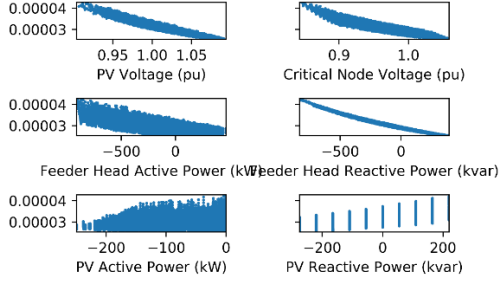

Fig. 4. $\delta^Q$ variation with respect to other measurements

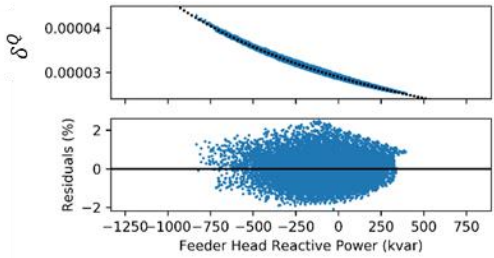

Fig. 5. $\delta^Q$ with curve fit and residuals

The voltage-sensitivity estimator also accounts for the impact of tap changes on voltage variations. Because the voltage sensitivity with respect to tap changes does not vary significantly across different scenarios, the voltage change caused by a tap change is only calculated for one scenario and ignoring the reactive power injection/absorption from the PV plants. The voltage sensitivity, $\delta_{i,r}^{VR}$, is calculated as

$$\delta_{i,r}^{VR} = V_i^r - V_i^0 \quad \forall r \in R, \forall i \in N \quad (4)$$

where $V_i^r$ is the voltage at node $i$ after a single tap change on regulator $r$ and $V_i^0$ is the voltage before the tap change.

*E. Online Operation*

For online operation, we first use the regression models along with limited measurements to calculate the critical voltages and sensitivities. Then, those values are input to the CVVC optimization to determine the optimal control actions.

*1) Voltage Estimation*

The voltage at each critical node is calculated as

$$\hat{V}_i = \sum_{m \in M} \alpha_{m,i} m + \sum_{r \in R} \delta_{i,r}^{VR}(T_r^t - T_r^0) \quad (5)$$

where $M$ is the set of available active and reactive power measurements at the feeder head and PV plant locations and $\alpha_{m,i}$ represents the corresponding regression coefficients for estimating voltage at node $i$. The first term of equation (5) estimates the voltage at node $i$ assuming the regulators are each on the same tap as they were during the initial training power flows. The second term superimposes the voltage change caused by the difference in regulator tap positions compared to those used in the training power flows using the voltage sensitivity calculated by (4). Thus, the CVVC controller can determine the voltages at all critical nodes using (5) with only active and reactive power measurements at the feeder head and PV plant locations and the current tap positions of each regulator.

Next, the voltage sensitivities are calculated based on the reactive power measurement at the substation and the fitted $\delta^Q$ curves. The $\delta^Q$ values are calculated for each critical node and for each PV plant. The critical node voltages and voltage sensitivities are input into the CVVC optimization along with the PV and VR device constraints.

*2) Voltage Estimation Correction*

Voltage estimation error, $\varepsilon_r$, is introduced because nodal voltages will vary when loads and PVs are distributed differently throughout the feeder. The use of static $\delta_{i,r}^{VR}$ values will also introduce a small error in calculating nodal voltage changes. Voltage measurements at the secondary bus of voltage regulator, $V_r$, are normally used in conventional local regulator control. Therefore, $\varepsilon_r$ is calculated as

$$\varepsilon_r = V_r - \hat{V}_r \quad \forall r \in R \quad (6)$$

where $\hat{V}_r$ is the estimated voltage at the secondary bus of each voltage regulator. Thus, (5) becomes

$$\hat{V}_i = \sum_{m \in M} \alpha_{m,i} m + \sum_{r \in R} \delta_{i,r}^{VR}(T_r^t - T_r^0) + \varepsilon_{r_i} \quad (7)$$

where $r_i$ is the regulator immediately upstream of node $i$. Thus, we assume that voltages downstream of a regulator will have similar voltage estimation errors as the regulator bus due to their close electrical connection as illustrated in Fig. 6.

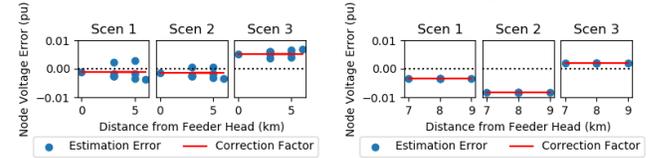

Fig. 6. Voltage error and correction factor, $\varepsilon_r$, for nodes downstream of two VRs for 3 different scenarios

| Algorithm 2: Voltage Estimation with Measurement Correction |
| --- |
| 1: Measure real and reactive power at feeder head and PV plants |
| 2: Measure voltage at each regulator secondary, $V_r^{meas}$ |
| 3: Estimate voltages at each regulator secondary, $V_r^{est}$, using (5) |
| 4: Calculate correction factor, $V_r^{corr}$, using (6) |
| 5: Estimate voltage at each critical node using (7) |

*3) Command Implementation*

Once the optimal commands are decided, the reactive power set points can be sent directly to the smart inverters to be implemented. For voltage regulator commands, the optimal tap positions cannot be implemented directly using conventional regulator controllers. Conventional regulator controllers operate based on a voltage set point, $V_T$, with a dead band, $B$, and time delay. Therefore, the desired tap position, $T$, needs to be

translated into $V_T$, which can be implemented by the regulator controller by

$$V_T = \begin{cases} V_s^1 + \left(\frac{1}{2}B - \frac{1}{2}T\right) & \text{if } dT > 0 \\ V_s^1 - \left(\frac{1}{2}B - \frac{1}{2}T\right) & \text{if } dT < 0 \\ V_s^1 & \text{if } dT = 0 \end{cases} \quad (8)$$

If no tap change is needed, the set point is set to the expected voltage at the regulator's secondary bus, $V_s^1$, after all other control action are implemented, which is calculated using the voltage sensitivities. This ensures that the regulator will be at the center of its band when the command is implemented and there will be no tap changes unless a voltage change with magnitude at least half of the VR's band occurs. If a tap change is needed, the set point is chosen such that the desired tap position will be the first tap within the regulator's new band. This forces the regulator to change taps appropriately to reach the new band.

Implementing tap changes by adjusting the local regulator set point achieves three benefits. First, the existing local regulator controller can be used to implement centrally optimized commands. Second, the local regulator control can operate to prevent over-voltages and under-voltages that may occur if the voltage continues to increase or decrease throughout the control interval. Finally, if there is some error in the voltage estimate, using the local controller to achieve a desired voltage set point ensures that the nodes near the regulator secondary will achieve the expected voltage despite estimation errors.

## III. OPTIMIZATION FORMULATION

The CVVC optimization determines the best combination of reactive power control at each PV plant and tap changes for each regulator. Therefore, the decision variables are the change in tap position up $\Delta T_r^+$ or down, $\Delta T_r^-$ for each regulator $r \in R$ and reactive power injection $Q_k^+$ and absorption $Q_k^-$ by each PV plant $k \in K$.

The CVVC problem is formulated as

$$\min \left( \sum C_r(\Delta T_r^+, \Delta T_r^-) + \sum C_k(Q_k^+, Q_k^-) \right) \quad (9)$$

$$\underline{Q_k} < Q_k^+ - Q_k^- < \overline{Q_k} \quad \forall k \in K \quad (10)$$

$$\underline{T_r} < T_r + \Delta T_r^+ - \Delta T_r^- < \overline{T_r} \quad \forall r \in R \quad (11)$$

$$\Delta T_r^+ + \Delta T_r^- < \overline{\Delta T} \quad \forall r \in R \quad (12)$$

$$\underline{V} < V_i^0 + (Q_k^+ - Q_k - Q_k^0) \times \delta_{i,k}^Q +$$
$$(\Delta T_r^+ - \Delta T_r^-) \times \delta_{i,r}^{VR} < \overline{V} \quad \forall i \in N \quad (13)$$

$$C_r(\Delta T_r^+, \Delta T_r^-) = c_r \times (\Delta T_r^+ + \Delta T_r^-) \quad \forall r \in R \quad (14)$$

$$C_k(Q_k^+, Q_k^-) = c_k \times (Q_k^+ + Q_k^-) \quad \forall i \in I \quad (15)$$

The objective of the CVVC optimization is to minimize the cost of control actions. (10) gives the reactive power constraints for each PV plant considering the total capacity and the current active power output. (11) constrains the resulting tap position within the voltage regulator's limits. (12) limits each voltage regulator to a reasonable number of tap changes for each 5-minute control interval. (13) enforces the voltage constraints at each critical node, using the voltage sensitivity matrices to calculate the expected voltages after control actions are implemented. (14) and (15) give the cost functions for tap changes and PV reactive power, respectively. Note that the cost functions are simply flat rates for each tap change and each kvar of reactive power injected or absorbed.

In some cases, the voltages cannot be fully corrected by the available resources. Therefore, an additional set of decision variables $V_i^{vio}$ is added for each critical node $i \in N$ to allow small voltage violations. The voltage constraint in (13) is replaced by (16) which allows for controlled voltage violations at each node. An updated objective function given by (17) uses the sum of squares of $V_i^{vio}$ as a penalty in the objective. The optimization is solved first with equations (9) through (15) to enforce all voltage constraints.

In cases where the problem is infeasible, the controller will minimize the sum of squared voltage violations instead of enforcing voltage constraints. So (9) is replaced by

$$\min \alpha \times \left( \sum C_r(\Delta T_r^+, \Delta T_r^-) + \sum C_k(Q_k^+, Q_k^-) \right) +$$
$$\beta \times \sum_N V_i^{vio^2}$$
$$\underline{V} - V_i^{vio} < V_i^0 + (Q_k^+ - Q_k - Q_k^0) \times \delta_{i,k}^Q +$$
$$(\Delta T_r^+ - \Delta T_r^-) \times \delta_{i,r}^{VR} < \overline{V} + V_n^{vio} \quad \forall i \in N \quad (17)$$

## IV. SIMULATION RESULTS

### A. Test Feeders and Data Sets

The estimator is tested in conjunction with the CVVC on a real, unbalanced rural circuit. The rural circuit has 1003 buses (1284 single-phase nodes) and includes 2 three-phase regulators and 2 single-phase regulators. There are 3 PV plants added to each circuit to give a total PV penetration equal to 100% of peak load as shown in Fig. 7a. One-year, one-minute resolution load profiles for each load node were allocated using the algorithm presented in [13] to account for load diversity. One-year, one-minute resolution PV profiles are taken from Pecan street data [14] and scaled so the peak PV output matches the PV capacity for each plant.

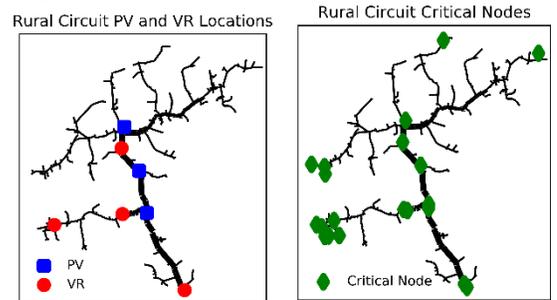

Fig. 7. Test feeder map with (a) PV and VR locations and (b) critical nodes

### B. Case Setup

To assess the efficacy of the proposed methods, four cases are compared as summarized in Table I.

TABLE I. SIMULATION CASE SETUP

| Case | Voltages Used | Voltages | Voltage Sensitivities | VR control |
|---|---|---|---|---|
| A | All Nodes | Measured | Perturbation | Tap positions |
| B | Critical Nodes | Estimator | Estimator | Tap positions |

| C | Critical Nodes | Estimator | Estimator | Set points |
| D | Critical Nodes | Estimator/correction | Estimator | Set points |

In case A, the CVVC is run assuming the controller has full visibility of the entire feeder. Every 5 minutes, the voltage inputs are taken directly from power flow results. The voltage sensitivities are calculated at each time step using the perturbation method as in [8]. The optimization formulation is the same as discussed in section IV except that, in the full visibility case, the voltage constraints are considered at every node instead of only critical nodes. The resulting optimal control actions are implemented for the next 5 minutes of simulation, and tap positions are implemented directly without translation to voltage set points.

In case B, the CVVC controller will use the critical-nodes based volt-var optimization algorithm to determine the optimal tap changes for voltage regulators. Voltages at the critical nodes are estimated by voltage and voltage sensitivity estimators with no measurement correction. The one-year load and PV profiles are used as the basis for creating the representative scenarios as discussed in section IIA. The estimator is trained according to section II. Every 5 minutes, the power measurements are taken and used to estimate the critical node voltages and sensitivities. Those values are input into the CVVC optimization and the actions are implemented as in case A.

Case C builds on case B but instead of implementing tap positions directly, the optimal tap changes are translated into voltage set points. Each regulator is given a bandwidth of 4V on a 120V base.

Case D builds on case C but with voltage correction implemented using measurements at each VR secondary bus. Additional simulations were run to illustrate the impact of different bandwidths for cases C and D.

For each case, the circuit is simulated in OpenDSS for 4 months (i.e., 1 month in each season) at one-minute resolution. For all cases, the voltage limits in the optimization were set to (0.975,1.04) to provide a 0.01 margin for error within the actual desired limits of (0.965,1.05).

### C. Critical Nodes Selected

47 critical nodes are selected as shown in Fig. 7b, representing 3.7% of all nodes.

### D. Efficacy of Representative Scenario Selection

The PV-Load-center based representative scenario selection method proposed in section IIB is compared to the random selection approach, in which six scenarios are selected randomly from each load and PV block.

The voltage estimator is first trained on the rural circuit and then used to estimate voltages throughout the winter month. As shown in Fig. 8a, the guided selection method reduces the bias in voltage estimates by ensuring that the full range of load and PV distributions is considered. With random selection, more common load and PV distributions are more likely to be selected which leads to bias and higher error in less common scenarios.

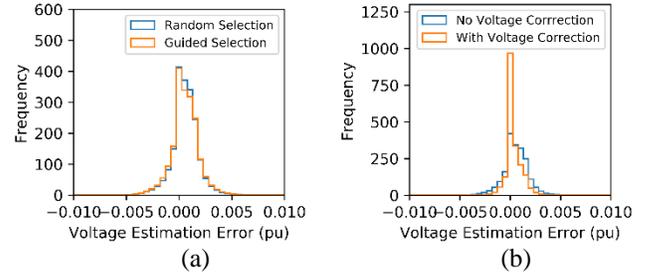

Fig. 8. Voltage estimation error with (a) random and guided selection and (b) with and without voltage correction

### E. Efficacy of Voltage Measurement Correction

In Fig. 8b, we compared voltage estimation errors for cases with and without voltage correction (i.e., using VR secondary voltage measurements for voltage correction) by running the rural circuit model for one month in winter. For both methods, voltage estimation errors are within [-0.01, 0.01] p.u., which is adequate for volt-var control. However, the estimates are much more accurate with less bias when voltage estimation errors are corrected by measurements.

### F. Overall Performance Comparison

Six performance metrics are used to quantitatively compare the performance of the four CVVC algorithms. To compare the voltage regulation effects, we compute the maximum voltage violation magnitude (MVVM) and the number of voltage violations (NVV) based on 10-minute rolling average voltage measurements at every node in the system. To compare the control resource usage, we calculate the number of tap changes (NTC), the reactive power (Q) cost, the tap change (T) cost, and total cost for each month of control. Table II summarizes the simulation results.

TABLE II. RURAL CIRCUIT RESULTS SUMMARY

| Season | Case | MVVM | NVV | NTC | Q Cost | T Cost | Total Cost |
|---|---|---|---|---|---|---|---|
| Winter month | A | 0.0072 | 39 | 411 | 694.30 | 57.54 | 751.84 |
| | B | 0.0007 | 43 | 456 | 605.19 | 63.84 | 669.03 |
| | C | 0 | 0 | 514 | 582.62 | 71.96 | 654.98 |
| | D | 0 | 0 | 665 | 591.61 | 93.10 | 684.71 |
| Spring month | A | 0 | 0 | 39 | 361.32 | 5.46 | 366.78 |
| | B | 0 | 0 | 21 | 451.24 | 2.94 | 454.18 |
| | C | 0 | 0 | 72 | 535.55 | 10.08 | 545.63 |
| | D | 0 | 0 | 77 | 482.08 | 10.78 | 492.86 |
| Summer month | A | 0 | 0 | 266 | 562.31 | 37.24 | 599.55 |
| | B | 0 | 0 | 251 | 513.52 | 35.14 | 548.66 |
| | C | 0.0007 | 98 | 437 | 552.23 | 61.18 | 613.41 |
| | D | 0 | 0 | 449 | 527.34 | 62.86 | 590.20 |
| Fall month | A | 0 | 0 | 11 | 269.27 | 1.54 | 270.81 |
| | B | 0 | 0 | 18 | 154.66 | 2.52 | 157.18 |
| | C | 0 | 0 | 37 | 221.31 | 6.44 | 227.75 |
| | D | 0 | 0 | 51 | 253.15 | 7.14 | 260.29 |

From the results, we have the following observations

*1) Impact of Voltage Estimation and Use of Critical Nodes*

Performance indicators match closely between cases A and B in all seasons, which shows that the estimator can provide adequate voltage visibility and sensitivity estimates to the

CVVC to provide a similar level of voltage control at similar cost compared to a system with full visibility. Using only critical nodes in the optimization does not lead to voltage violations at unmonitored nodes in these cases.

*2) Impact of VR set points*

The use of VR set points rather than tap positions in cased C and D invariably increases the total number of tap changes. However, this change is necessary to implement VR actions using existing VR controllers. The addition of local control by VRs through the use of set points also eliminates the small number of voltage violations observed in the winter month in cases A and B. Those violations occur due to voltage changes within a control interval, and can be corrected by the VR's local control.

*3) Impact of voltage correction*

The addition of voltage correction in case D eliminates the 98 voltage violations observed in the summer month in case C. In all cases, the addition of voltage corrections in case D brings the total control cost closer to the observed value with full visibility in case A.

### G. Impact of Voltage Regulator Deadband Settings

Table III compares the results with different VR dead bands for cases C and D on the rural circuit. Increasing the VR bandwidth is one way to reduce the number of unnecessary tap changes when local control is used [1]. The results in table III show that this holds true when centrally optimized VR voltage set points are used. When the bandwidth is reduced to 2V, excessive tap changes are observed. In addition, voltage violations become common because the VRs tend to operate on small local voltage fluctuations to the detriment of the overall system.

TABLE III. SUMMARY OF VR BANDWIDTH IMPACTS ON THE RURAL CIRCUIT IN WINTER

| Regulator Dead Band | Case | MVVM | NVV | NTC | Q Cost | T Cost | Total Cost |
|---|---|---|---|---|---|---|---|
| 4V | C | 0 | 0 | 514 | 582.62 | 71.96 | 654.98 |
|    | D | 0 | 0 | 665 | 591.61 | 93.10 | 684.71 |
| 3V | C | 0 | 0 | 650 | 597.14 | 91.00 | 688.14 |
|    | D | 0.0009 | 29 | 710 | 600.60 | 99.40 | 700.00 |
| 2V | C | 0.0047 | 172 | 1445 | 875.59 | 202.30 | 1077.90 |
|    | D | 0.0048 | 241 | 1354 | 980.43 | 189.56 | 1169.99 |

### H. Comparison Between Case A and Case D

We compare the performance of the proposed regression-based method for estimating voltages and voltage sensitivities for volt-var control with voltage correction with the CVVC with full visibility, where real and reactive power measurements of all nodes are used. The nodal voltage magnitudes, reactive power injections, and tap positions for cases A and D for the rural circuit in winter are compared in Figs. 9, 10, and 11, respectively. Tables V and VI summarize the actions taken by each PV and VR, respectively.

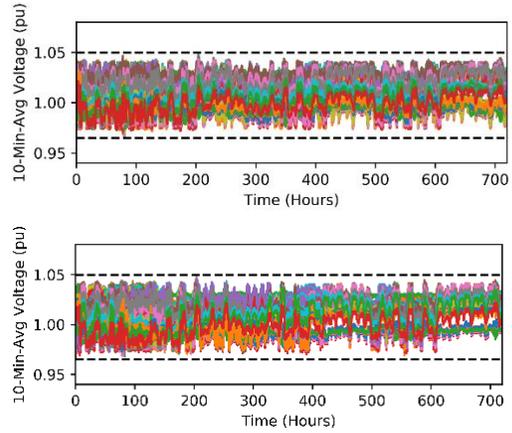

Fig. 9. 10-minute average voltage results on the rural curcuit in winter

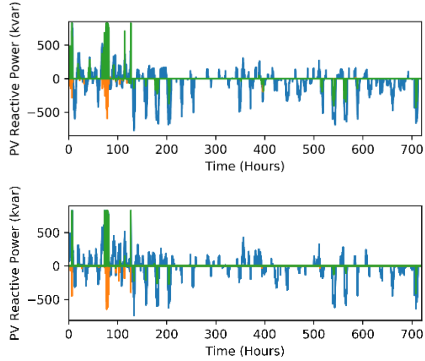

Fig. 10. Reactive power injections by each PV plant on the rural circuit in winter

TABLE IV. PV REACTIVE POWER CONTRIBUTIONS IN WINTER

|    | Reactive Power (Mvar) | |
|---|---|---|
| PV | Case A | Case D |
| 1 | 713.924 | 638.121 |
| 2 | 20.412 | 18.406 |
| 3 | 98.821 | 53.409 |

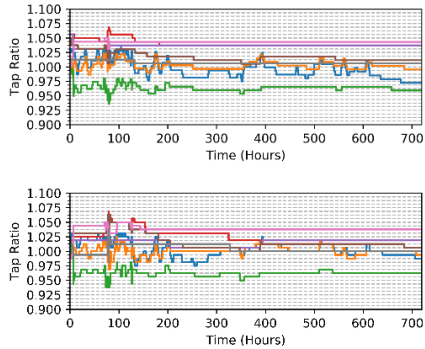

Fig. 11. Tap positions for each VR on the rural circuit in winter

TABLE V. VR TAP CHANGES IN WINTER

|    | Tap Changes | |
|---|---|---|
| VR | Case A | Case D |
| 1a | 179 | 148 |
| 1b | 95 | 179 |
| 1c | 75 | 113 |
| 2a | 9 | 40 |
| 3a | 9 | 24 |
| 3b | 24 | 74 |
| 3c | 18 | 46 |
| 4a | 2 | 41 |

The following observations are made:
- For both approaches, the nodal voltages remain within the desired limits in nearly all cases.
- Reactive power control at each PV plant is very similar between the two cases, which shows that the CVVC can accurately assign reactive power commands to the most effective PV plants based on the estimated voltages and sensitivities using measurements from only the feeder head and the PV sites corrected by measurements from voltage regulators.
- The tap positions for each regulator are similar under both CVVC control schemes. The number of tap changes increases for nearly all regulators in case D compared to case A. These are mainly triggered by the local regulator controller between optimizations.

## V. Conclusion

This paper addresses several practical challenges of implementing CVVC on distribution feeders. First, a method to estimate voltages and voltage sensitivities online using regressions on offline power flow results is presented. The estimator provides visibility of the critical node voltages across the distribution feeder using only active and reactive power measurements at the substation and at each PV plant. This estimator provides the real-time data needed to run a centralized volt-var control algorithm using very limited communication infrastructure, which greatly reduces the capital needed to implement a centralized volt-var control system. By identifying critical nodes, the number of constraints is greatly reduced compared to the full visibility case, which improves online computational time. Finally, the implementation of optimal tap positions using conventional regulator control equipment is addressed. The proposed method is effective on large circuits with diverse loads and multiple large PV plants with diverse output profiles.

This method assumes an accurate circuit model, reliable measurements, and reliable communication to the feeder head and PV plants. The voltage correction mechanism proposed here can mitigate some impacts of modeling and measurement errors. However, additional work is needed to understand the robustness of this method to circuit modeling errors, measurement errors and communication failures.

## References

[1] W. Ren and H. Ghassempouraghamolki, "Tuning of Voltage Regulator Control in Distribution," in IEEE Power and Energy Society General Meeting, Portland, 2018.

[2] Q. Zhou and J. W. Bialek, "Simplified Calculation of Voltage and Loss Sensitivity Factors in Distribution Networks," in 16th Power Systems Computation Conference (PSCC 2008), Glasgow, Scotland, 2008.

[3] X. Zhu, J. Wang, D. Mulcahy, D. L. Lubkeman, N. Lu, N. Samaan and R. Huang, "Voltage-Load Sensitivity Matrix Based Demand Response for voltage Control in High solar Penetration Distribution Feeder," in IEEE Power & Energy Society General Meeting, Chicago, 2017.

[4] K. Christakou, J.-Y. LeBoudec, M. Paolone and D.-C. Tomozei, "Efficient computation of Sensitivity Coefficients of Node Voltages and Line Currents in Unbalanced Radial Electrical Distribution Networks," IEEE Transactions on Smart Grid, vol. 4, no. 2, pp. 741-750, 2013.

[5] K. Christakou, D.-C. Tomozei, J.-Y. Le Boudec and M. Paolone, "GECN: Primary Voltage Control for Active Distribution Networks via Real-time Demand-Response," IEEE Transactions on Smart Grid, vol. 5, no. 2, pp. 662-631, 2014.

[6] R. A. Jabr and I. Džafić, "Sensitivity-Based Discrete Coordinate-Descent for Volt/VAr Control in Distribution Networks," IEEE Transactions on Power Systems, vol. 31, no. 6, pp. 4670 - 4678, 2016.

[7] C. McEntee, D. Mulcahy, J. Wang, X. Zhu and N. Lu, "A VSM-Based DER Dispatch MINLP for Volt-VAR Control in Unbalanced Power Distribution Systems," in IEEE Power & Energy Society General Meeting, Atlanta, GA, 2019.

[8] Z. Zhang, L. F. Ochoa and G. Valverde, "A novel Voltage Sensitivity Approach for the Decentralized Control of DG Plants," IEEE Transactions on Power Systems, vol. 33, no. 2, pp. 1566-1576, 2018.

[9] K. Dehghanpour, Z. Wang, J. Wang, Y. Yuan and F. Bu, "A survey on state estimation techniques and challenges in smart distribution systems," IEEE Transactions on Smart Grid, vol. 10, no. 2, pp. 2312-2322, 2019.

[10] S. Weckx, R. D'Hulst and J. Driesen, "Voltage Sensitivity Analysis of a Laboratory Distribution Grid with Incomplete Data," IEEE Transactions on Smart Grid, vol. 6, no. 3, pp. 1271-1280, 2015.

[11] J. Deboever, S. Grivalja, J. Peppanen, M. Rylander and J. Smith, "Practical data-driven methods to improve the accuracy and detail of hosting capacity analysis," in IEEE 7th World Conference on Photovoltaic Energy Conversion, Waikoloa Village, Hawaii, USA, 2018.

[12] V. Rigoni, A. Soroudi and A. Keane, "Use of fitted polynomials for the decentralised estimation of network variables in unbalanced radial LV feeders," IET Generation, Transmission & Distribution, vol. 14, no. 12, pp. 2368-2377, 2020.

[13] J. Wang, X. Zhu, D. Mulcahy, C. McEntee, D. Lubkeman, N. Lu, N. Samaan, B. Werts and A. Kling, "A Two-Step Load Disaggregation Algorithm for Quasi-static Time-series Analysis on Actual Distribution Feeders," in IEEE Power and Energy Society General Meeting, Portland, 2018.

[14] "Residential load data collected by PECAN STREET website: https://dataport.pecanstreet.org/data/interactive," [Online].